\documentclass[prd,10pt,aps,showpacs,twocolumn,longbibliography,nofootinbib,amsmath,amssymb,superscriptaddress,longbibliography]{revtex4-1}
\usepackage[colorlinks=True, linkcolor=violet, citecolor=teal, urlcolor=blue]{hyperref}

\usepackage[T1]{fontenc}
\usepackage[english]{babel}
\usepackage{amsmath,amssymb,amsfonts}
\usepackage{graphicx}
\usepackage{cancel}
\usepackage{braket}
\usepackage{xcolor}
\usepackage{bookmark}
\usepackage{chngcntr}
\usepackage{nicefrac}
\usepackage{combelow}

\usepackage{comment}
\usepackage{bm}
\usepackage{bbm}
\usepackage{braket}
\usepackage{slashed}

\newcommand{\avr}[1]{{\left\langle #1 \right\rangle}}

\definecolor{purple}{rgb}{0.8,0,0.6}
\newcommand{\corr}[1]{\textcolor{black}{#1}}

\allowdisplaybreaks

\begin{document}
\title{Vortex wall phase in fractonic XY-plaquette model on square lattice}

\author{A. M. Begun}
\affiliation{Nordita, Stockholm University, Roslagstullsbacken 23, SE-106 91 Stockholm, Sweden}
\affiliation{Pacific Quantum Center, Far Eastern Federal University, \corr{690922} Vladivostok, Russia}

\author{M. N. Chernodub}
\affiliation{Institut Denis Poisson UMR 7013, Universit\'e de Tours, 37200 Tours, France}
\affiliation{Department of Physics, West University of Timi\cb{s}oara,  Bd.~Vasile P\^arvan 4, Timi\cb{s}oara 300223, Romania}

\author{V. A. Goy}
\affiliation{Pacific Quantum Center, Far Eastern Federal University, \corr{690922} Vladivostok, Russia}
\affiliation{\corr{Institute of Automation and Control Processes, Far Eastern Branch, Russian Academy of Science, 5 Radio Str., Vladivostok 690041, Russia}}

\author{A. V. Molochkov}
\affiliation{Beijing Institute of Mathematical Sciences and Applications, Tsinghua University, 101408, Beijing, China}
\affiliation{Institut Denis Poisson UMR 7013, Universit\'e de Tours, 37200 Tours, France}

\begin{abstract}
The XY-plaquette model is the most straightforward lattice realization of a broad class of fractonic field theories that host quasiparticles with restricted mobility. The plaquette interaction appears naturally as a ring-exchange term in the low-energy description of exciton Bose liquids, cold atomic gases, and quantum dimer models. Using first-principle Monte Carlo simulations, we study the phase diagram and the vortex dynamics in the XY-plaquette model on a square lattice in two spatial dimensions. In its minimal formulation, the model contains a ring-exchange plaquette term in two spatial dimensions and a standard XY-link term in the (imaginary) time direction. We show that the phase diagram of the minimal XY-plaquette model possesses two phases: (i) a disordered vortex-dominated phase in which a single percolating vortex trajectory occupies the whole 3d spacetime; (ii) a partially disordered phase in which the vortices become partially immobile, with their worldlines strictly confined to one or several infinite two-dimensional planes. The spatial positions and spatial orientations (along $x$ or $y$ axis) of these vortex domain walls appear to be spontaneous. Individual vortices form a disordered system within each vortex domain wall, so the fractal spacetime dimension of vortex trajectories approaches $D_f = 2$. We argue that the appearance of the vortex walls could be interpreted as a consequence of the spontaneous breaking of a global internal symmetry in the compact XY-plaquette model.
\end{abstract}
%
\date{\today}
\maketitle
\section{Introduction}

Fractons constitute an exotic class of quasiparticles, which exhibit a unique deviation from traditional particle behavior: while an isolated fracton does not move under an applied force, a set of fractons can propagate freely provided their total dipole moment (or its multipole generalizations) remains conserved~\cite{Chamon:2004lew, Haah:2011drr, Vijay:2015mka, Vijay:2016phm, Pretko:2016kxt, Pretko2018}. In specific systems, fractons can exhibit mobility through the formation of composite bound states that can be either stable or mobile, while in other systems, the fractons disintegrate directly into the vacuum~\cite{Vijay:2016phm}. Additionally, an isolated fracton can achieve motion at the expense of generating new fractons from the vacuum with each movement step. Nevertheless, without a continuous energy supply to facilitate this process of particle creation, a single fracton remains stationary. Recent reviews of the subject of the rapidly developing theories of fractons, including the unusual symmetries of the underlying field theories and emerging fracton hydrodynamics, can be found in~\cite{Nandkishore2019, Pretko:2020cko, Grosvenor:2021hkn, Marsot:2022imf, Gromov:2022cxa}.

In field theoretical language, the properties of fracton excitations can be understood via multipolar symmetries and corresponding conservation laws~\cite{Pretko:2016kxt}. One of the simplest examples of such models is given, in three spacetime dimensions, by the action~\cite{Pretko2018}:
\begin{align}
S = \int d^3 x \biggl[
\frac{\mu_t}{2}  \bigl(\partial_{t} \phi\bigr)^2
- \frac{1}{2\mu_{xy}} \bigl(\partial_{x} \partial_{y} \phi\bigr)^2\biggr]\,,
\label{eq_L_continuum_rescaled}
\end{align}
\corr{where $\mu_t$ and $\mu_{xy}$ are parameters with the dimension of mass implying that the field $\phi$ is dimensionless. Action~\eqref{eq_L_continuum_rescaled} is written in Lorentzian signature with the spacetime coordinate $\boldsymbol{x} = (t,x,y)$.} Below, we use Euclidean spacetime notations suitable for numerical simulations implemented in our paper. \corr{We also use periodic boundary conditions for all coordinates $t$, $x$ and $y$.}

The Hamiltonian of the model~\eqref{eq_L_continuum_rescaled} is invariant under an infinitesimal polynomial shift symmetry~\cite{Gromov:2022cxa}:
\begin{align}
\delta \phi = c_0 + c_x x + c_y y\,,
\label{eq_gauge_transformation}
\end{align}
with three arbitrary real-valued parameters: $c_0$, $c_x$ and $c_y$. 
\corr{
The infinitesimal gauge transformations correspond to a momentum dipole symmetry~\cite{You2021}. They can be promoted to the full momentum dipole symmetry, which shifts the scalar field by arbitrary functions, $f(x) $ and $ g(y)$, that depend on one spatial coordinate only:
\begin{align}
	\phi(x,y,t) \to \phi(x,y,t) + f(x) + g(y)\,,
    \label{eq_gauge_transformation2}
\end{align}
}

\corr{The invariance under transformations~\eqref{eq_gauge_transformation} or \eqref{eq_gauge_transformation2}, leads, according to Noether's theorem, to the conservation of total charge and total dipole moment of the fields. In particular, the classical equation of the model~\eqref{eq_L_continuum_rescaled},
\begin{align}
	\mu_t \partial_{t}^2 \phi + \frac{1}{\mu_{xy}} \partial^2_{x} \partial^2_{y} \phi = 0\,,
\end{align}
gives rise to a global momentum dipole symmetry associated with the currents~\cite{You2021} ({\it cf.} also a review section of Ref.~\cite{Seiberg:2020bhn}):
\begin{align}
	J^t = \mu_t \partial^t \phi\,, \qquad
	J^{xy} = - \frac{1}{\mu_{xy}} \partial^x \partial^y \phi\,,
\end{align}
that are subjected to the conservation law:
\begin{align}
    \partial_t J^t - \partial_{x} \partial_{y} J^{xy} = 0\,.
\end{align}
}

\corr{The conserved charges of the momentum dipole symmetry can be given in terms of the line integrals along the straight paths, parallel to $x$ or $y$ axis, and closed due to the periodic boundary conditions:
\begin{align}
	Q^x(x) = \oint d y J^t\,, \qquad
	Q^y(y) = \oint d x J^t\,.
\end{align}
Below, we will consider a compact lattice realization of the model~\eqref{eq_L_continuum_rescaled} in which the field $\phi$ is a $2\pi$-period field. 
}

Generally, due to the conservation of the dipole moment, a single particle cannot move, while multi-particle ensembles can propagate, provided the value of their total dipole moment remains unaltered~\cite{Pretko:2016kxt}. The model~\eqref{eq_L_continuum_rescaled} and its generalizations have various applications in condensed matter physics~\cite{Grosvenor:2021hkn, Stahl:2021sgi, Lake:2022ico, Gorantla:2022eem, Radzihovsky:2022fnr}.

Theories of uncharged scalar fields, similar to the model~\eqref{eq_L_continuum_rescaled}, are usually associated with superfluid systems where the scalar field condenses. The condensate can contain topological defects related to the scalar field known as superfluid vortices. In our paper, we are interested in a straightforward $XY$-type generalization of the continuum model~\eqref{eq_L_continuum_rescaled} ---which is the simplest lattice representative of the class of the fractonic field theories--- that allows us to include topological vortices. To this end, it is sufficient to treat the field $\phi_0$ as a phase of a complex scalar condensate. In other words, we "compactify" this simplest fractonic model and study the vortices associated with the winding singularities in the compact scalar field.

Such an approach works in ordinary superfluids within a widely used approximation of uniform superfluid density. In this case, the radial part of the bosonic superfluid condensate is fixed, and the system possesses only one dynamical field: the (necessarily compact) phase of the superfluid condensate. The discretized (lattice) version of a superfluid model is the celebrated XY model~\cite{Itzykson1989}, which describes one scalar degree of freedom per each lattice site coupled to each other via a nearest-neighbor interaction. 

However, even in the XY model, the properties of the vortex excitations are far from being simple. The planar, two-dimensional version of the model experiences the Berezinskii–Kosterlitz–Thouless (BKT) phase transition~\cite{Berezinsky:1970fr, Berezinsky:1972rfj, Kosterlitz1973} associated with dipole-binding of the vortex-antivortex pairs at low temperatures to unpaired vortices and anti-vortices at a critical BKT temperature. At three spatial dimensions, the studies of the model involve significant numerical efforts~\cite{Chester:2019ifh, Hasenbusch2019}. Therefore, we implement a lattice regularization of the compact-field version of the fractonic field theory~\eqref{eq_L_continuum_rescaled} and use numerical simulations to explore its discretized XY-like formulation. 

The lattice discretization of the compact version of the system~\eqref{eq_L_continuum_rescaled} is described by a plaquette-extended XY model~\cite{Paramekanti2002}. \corr{A nontrivial phase structure of this model has been observed~\cite{Paramekanti2002}.} This model and its extensions, which we discuss in detail in the next section, have been studied intensively in the Villain formulation of the action \corr{(both in original and modified extensions)}, which is proven to be very convenient for analytical approaches to the problem~\cite{Gorantla:2020xap, Seiberg:2020bhn, Seiberg:2020wsg, Fazza:2022fss}. The theoretical analysis indicates that topological defects, similarly to the original fracton fields, can exhibit restricted mobility~\cite{Gorantla:2020xap, Giergiel2022}. For example, vortices may also follow a fractonic behavior by conserving the total dipole moment as well as a trace of the quadrupole moment of vorticity~\cite{Doshi2021}.

Our paper addresses several interconnected questions related to topological excitations in the fracton field models: What are the consequences of the restricted mobility of the excitations represented by the scalar field on the mobility of vortex singularities in this field? Do individual vortices move, or can they only propagate in the form of neutral dipole pairs? What is the phase diagram of the plaquette-XY model, which is the compact lattice generalization of the continuum model~\eqref{eq_L_continuum_rescaled}? If the latter is nontrivial, what are the properties of the vortices in the different phases? 

The structure of the paper is as follows. The lattice action, its naive continuum limit, and the definition of the vortices are described in Section~\ref{sec_lattice}. The results of the numerical simulations of the thermodynamics of the model and its phase diagram are given in Section~\ref{sec_results_phase}. The nature of the phase and the vortex properties are described in Section~\ref{sec_results_vortex}. Our conclusions are summarized in the last section. 

\section{The model and the vortices}
\label{sec_lattice}

\subsection{Lattice model}

We consider a plaquette generalization of the $XY$ model in the Euclidean three-dimensional spacetime~\cite{Paramekanti2002, Gorantla:2021svj, Lake2022}:
\begin{subequations}
\begin{align}
    S[\phi] & = S_\tau[\phi] + S_{xy}[\phi]\,,
\label{eq_S_phi}\\[1mm]
    S_\tau[\phi] & = \beta_\tau \sum_{\boldsymbol{x}} [1-\cos (\Delta_\tau \phi_{\boldsymbol{x}})]\,,
\label{eq_S_tau}\\ 
    S_{xy}[\phi] & = \beta_{xy} \sum_{\boldsymbol{x}} [1 - \cos (\Delta_x \Delta_y \phi_{\boldsymbol{x}})]\,,
\label{eq_S_xy}
\end{align}
\label{eq_S}
\end{subequations}
~\hskip -2mm where the single degree of freedom is represented by the compact U(1) scalar field $\phi_{\boldsymbol{x}}$ encoded in the phase of the site (matter) variable $U_{\boldsymbol{x}} = e^{i \phi_{\boldsymbol{x}}}$. In both terms of the action~\eqref{eq_S_phi}, the sum goes over the whole lattice ${\boldsymbol{x}} = (x,y,\tau)$, where $x$ and $y$ are the two-dimensional space coordinates and $\tau$ is the imaginary time variable. 

The notations in Eq.~\eqref{eq_S} are as follows. The first term in the action~\eqref{eq_S_tau} goes over temporal {\it links} $l_{{\boldsymbol{x}},\tau} = \{{\boldsymbol{x}},\tau\}$, 
\begin{align}
    \phi_{{\boldsymbol{x}},\tau} \equiv \Delta_\tau \phi_{\boldsymbol{x}} \equiv \phi_{\boldsymbol{x} + \hat{\boldsymbol{\tau}}} - \phi_{\boldsymbol{x}}\,, 
    \label{eq_d_phi_link}
\end{align}
where $\hat{\boldsymbol{\mu}}$ represents a unit (in lattice units) vector in the direction $\mu$. 

The second term in Eq.~\eqref{eq_S} is a lattice functional~\eqref{eq_S_xy} involving spatial {\it plaquettes}  $P_{{\boldsymbol{x}},xy} = \{{\boldsymbol{x}},xy\}$
\begin{align}
    \phi_{{\boldsymbol{x}},xy} \equiv \Delta_x \Delta_y \phi_{\boldsymbol{x}} & = \Delta_x (\phi_{\boldsymbol{x} + \hat{\boldsymbol{y}}} - \phi_{\boldsymbol{x}}) 
    \label{eq_d_phi_plaquette}\\
    & =  (\phi_{\boldsymbol{x} + \hat{\boldsymbol{x}} + \hat{\boldsymbol{y}}} - \phi_{\boldsymbol{x}+ \hat{\boldsymbol{x}}}) - (\phi_{\boldsymbol{x} + \hat{\boldsymbol{y}}} - \phi_{\boldsymbol{x}})\,. \nonumber 
\end{align}

Therefore, despite sharing a resemblance with the $XY$ model in the field content, the action of the system~\eqref{eq_S} is nowhere similar to the standard action of the XY model. The lattice functional~\eqref{eq_S} possesses two parameters\footnote{In the notations of Ref.~\cite{Gorantla:2021svj}, the identifications of the parameters are $\beta_\tau \equiv \beta_0$ and $\beta_{xy} \equiv \beta$.}: $\beta_\tau$ and $\beta_{xy}$, which control, respectively, the coupling between the two dimensional $xy$ planes via the term~\eqref{eq_S_tau} and the dynamics of the scalar $\phi_{\boldsymbol{x}}$ in each plane via the term~\eqref{eq_S_xy}. The couplings make the temporal $\tau$ and spatial, $x$ and $y$, coordinates different, but this difference is not the main reason to distinguish the lattice model~\eqref{eq_S} from the standard, albeit asymmetrical, XY model. The main distinction between the mentioned models appears in the different qualitative nature of the kinetic terms in the action terms \eqref{eq_S_tau} and \eqref{eq_S_xy}. 

\corr{We note that the Euclidean action~\eqref{eq_S} represents a specific lattice generalization of the original continuum model~\eqref{eq_L_continuum_rescaled}, which was formulated in Lorentzian spacetime. This lattice theory admits vortex excitations, the properties of which constitute one of the primary subjects of investigation in the present work. Alternatively, one could consider a compact lattice formulation based on a modified Villain representation, in which vortex excitations are suppressed~\cite{Gorantla:2021svj}. In this study, however, we focus on the formulation~\eqref{eq_S}, as it both supports vortex excitations and is well suited for numerical lattice simulations.}

The plaquette interaction~\eqref{eq_S_xy} in the model corresponds to the ring-exchange term, which has been extensively studied in the context of exciton Bose liquids~\cite{Paramekanti2002, Balents2003, Xu2007} and the theory of fractonic excitations~\cite{You2020, You:2021tmm, Giergiel2022}. By constructions, the ring exchange term represents a four-spin interaction where the phase differences between spins around a plaquette are crucial. Unlike the standard XY model, where the energy depends only on the phase difference between pairs of neighboring spins, the ring exchange term depends on the combined phase differences around the entire plaquette.

Before going further, it is noteworthy to highlight the symmetries of the action~\eqref{eq_S}. First of all, the model describes a {\it compact} scalar field. The compactness of the action is manifested in the discrete invariance 
\begin{align}
    \phi_{\boldsymbol{x}} \to \phi_{\boldsymbol{x}} + 2 \pi n_{\boldsymbol{x}}\,,
    \qquad 
    n_{\boldsymbol{x}} \in \mathbb{Z}\,,
    \label{eq_compactness}
\end{align}
under the discrete $\mathbb Z$ symmetry with an integer number $n_{\boldsymbol{x}}$ at every lattice site $\boldsymbol{x}$. The compactness of the model implies the existence of topological excitations called vortices. 
\corr{
Importantly, the lattice model also respects the global momentum dipole symmetry~\eqref{eq_gauge_transformation2} pertinent to the original model~\eqref{eq_L_continuum_rescaled}.
}

The lattice construction of vortices proceeds as follows~\cite{Savit1978, Shenoy1989, Gupta1992}. We identify the physical link variable,
\begin{align}
     {\bar \phi}_{{\boldsymbol{x}},\mu} 
     = {(\Delta \phi)}_{{\boldsymbol{x}},\mu} + 2 \pi k_{{\boldsymbol{x}},\mu} \in (-\pi,\pi]\,,
    \label{eq_bar_phi}
\end{align}
where the integer-valued link field $k_{{\boldsymbol{x}},\mu} \in \mathbb{Z}$ is chosen in such a way that the field~\eqref{eq_bar_phi} appears to be in the physical interval of values: $- \pi < {\bar \phi}_{{\boldsymbol{x}},\mu} \leqslant \pi$. Equation~\eqref{eq_bar_phi} is a generalization of the lattice derivative ${(\Delta \phi)}_{{\boldsymbol{x}},\mu}$ given in Eq.~\eqref{eq_d_phi_link} due to the compactness of the model~\eqref{eq_compactness}. 
\corr{Notice that the difference between the first derivative introduced in Eq.~\eqref{eq_d_phi_link} and the first derivative defined in Eq.~\eqref{eq_bar_phi} appears only in the additive terms ($2\pi$ multiplied by an integer number) which brings the finite difference to the canonical interval $- \pi < {\bar \phi}_{{\boldsymbol{x}},\mu} \leqslant \pi$.
}

The next step is to construct the vortex plaquette variable out of the physical link derivative~\eqref{eq_bar_phi}:
\begin{align}
    v_{{\boldsymbol{x},\mu\nu}} \equiv \frac{1}{2\pi}\left(
          {\bar \phi}_{{\boldsymbol{x}},\mu} 
        + {\bar \phi}_{{\boldsymbol{x} + \hat\mu},\nu} 
        - {\bar \phi}_{{\boldsymbol{x} + \hat\nu},\mu} 
        - {\bar \phi}_{{\boldsymbol{x}},\nu}\right)\,.
    \label{eq_v_P}
\end{align}
One can show that this quantity takes only integer values, $v_P \in \mathbb{Z}$. Moreover, the plaquette variable~\eqref{eq_v_P} has a conservation property, implying that a sum of the $v_P$ plaquettes over the sides $P$ of any elementary lattice cube $C$, taking into account their relative orientations, sums up to zero:
\begin{align}
    \sum_{P \in \partial C} (-1)^P v_P = 0\,.
    \label{eq_d_C}
\end{align}
To make the conservation property more visually appealing, it is convenient to identify a dual lattice that corresponds to the original lattice shifted by a half lattice spacing, $a/2$, in positive directions along all lattice axes. Then, the links ${}^* l$ of the dual lattice become normal to the plaquettes of the original lattice. Then, the set of plaquettes $v_P$ on the original lattice corresponds to a set of links ${}^* v_{{}^* l}$ on the dual lattice. The observation~\eqref{eq_d_C} implies that the vortex trajectories are closed, meaning that on the dual lattice, they form closed lines. The conservation condition on the original lattice~\eqref{eq_d_C} can be written in a compact form on the dual lattice:
\begin{align}
    \delta \,{}^*\! v = 0\,, \quad \Leftrightarrow \quad  
    \sum_{\,{}^*\! l \in {}^* s} (-1)^{{}^*\! l} v_{\,{}^*\! l} = 0\,,
    \label{eq_dual_conservation}
\end{align}
where the sum goes over the (six) dual lattice links ${}^*\! l$ (normal to the original lattice plaquettes $P$), which have the dual lattice site ${}^*\! s$ (a center of the original lattice cube $c$) as its beginning or end point, with $(-1)^{{}^*\! l} = +1$ and $(-1)^{{}^*\! l} = -1$ factors, respectively. 

In other words, the lattice quantity~\eqref{eq_v_P} has thus an interpretation of a winding number of a vortex that pierces a plaquette~\cite{Savit1978, Shenoy1989, Gupta1992}. As the vorticity number is conserved, the vortex that enters any volume (for example, an elementary lattice cube) should also quit this volume, implying that the sum over all vorticities at any closes surface is zero, hence~\eqref{eq_d_C}.

We associate the lattice density of the vorticity number, or the vortex vector density, ${\boldsymbol{v}} \equiv (v_\tau, v_x, v_y)$, with the corresponding plaquette variable: $v_i = (1/2) \epsilon_{ijk} v_{jk}$. It gives us
\begin{align}
    v_\tau = v_{xy}\,, \qquad v_{x} = v_{y \tau}, \qquad v_{y} = - v_{x\tau}\,.
    \label{eq_v_correspondence}
\end{align}
In our simulations, the overwhelming majority of links take values $v_i = 0, \pm 1$ $(i = 1,2,3)$. Hereafter, we remove the asterisk ``$*$'' from the vortex vector density $\boldsymbol{v}$ defined on the dual lattice to simplify the notation.

The lattice model~\eqref{eq_S} has been studied analytically in Ref.~\cite{Gorantla:2021svj} where the lattice action has been taken in the \corr{(modified)} Villain representation, which keeps all the symmetries of the model and facilitates theoretical calculations. \corr{One of the modification of the Villain model undertaken in Ref.~\cite{Gorantla:2021svj} was implemented to suppress vortices. In other words, no vortex-like fluxes appear in the spectrum in a result of the modification of the original formulation by Villain~\cite{Villain:1974ir}.} \corr{In our paper, we consider the original formulation of the model where the interactions between the fields are taken in the Wilson form and no special suppression of vortices has been implemented. Our numerical work can be considered as a natural generalization of  the analytical approach of Ref.~\cite{Gorantla:2021svj}.} 

We work with the Wilson-type of the lattice action~\eqref{eq_S} suitable for implementation in the numerical approach and concentrate on the properties of vortices that emerge in this emergent higher-rank field theory. It worth noticing that a class of compact models identical or similar to the one in question~\eqref{eq_S} are expected to feature fractonic dynamics of topological defects~\cite{You:2019cvs, You2022} and the phase structure~\cite{Grosvenor2023}.

\subsection{Continuum limit}

Does the lattice XY-like model~\eqref{eq_S} possess the continuum limit? The first term in its action~\eqref{eq_S} is a standard term that, in the continuum limit, ignoring the $2\pi$-vortex-like singularities, gives rise to the standard quadratic derivative, 
\begin{align}
    \beta_{\tau} \sum_{\boldsymbol{x}} [1 - \cos (\Delta_\tau \phi_{\boldsymbol{x}})] \to \frac{1}{2 g^2_\tau} \int d^3 x \, \bigl[\partial_\tau \phi({\boldsymbol{x}})\bigr]^2\,.
    \label{eq_continuum_1}
\end{align}
\corr{We stress that the last step in Eq.~\eqref{eq_continuum_1} assumes that the configuration is smooth implying the absence of vortices. Should vortices be present in the configuration, the resulting term should be supplemented with a singular contribution~\cite{Kleinert2008}.}

In the naive continuum limit ---assuming a smooth, no-vortex configuration--- one gets 
\begin{align}
\Delta_\mu \phi_{\boldsymbol{x}} = a^{3/2} \partial_\mu \phi(\boldsymbol{x}) + O(a^3)\,, 
\label{eq_phi_scaling_1}
\end{align}
where $a$ is a small lattice spacing, and the series in Eq.~\eqref{eq_d_phi_link} over $a$ are done with respect to the center of the lattice link. In Eq.~\eqref{eq_phi_scaling_1}, we have set the scaling $\phi_{\boldsymbol{x}} = a^{1/2} \phi(\boldsymbol{x})$, which relates the lattice scalar real field $\phi_{\boldsymbol{x}}$ with its continuum counterpart, $\phi(\boldsymbol{x})$.\footnote{Note that contrary to the dimensionless lattice field $\phi_{\boldsymbol{x}}$, the continuum field $\phi(\boldsymbol{x})$ has a dimension [mass]${}^{1/2}$.} Such a scaling is typical for scalar theories in a three-dimensional spacetime. Finally, we arrive at Eq.~\eqref{eq_continuum_1} in the leading order with the subleading $O(a^2)$ corrections vanishing in the continuum limit. For further convenience, we also redefined the coupling constant 
\begin{align}    
    \beta_{\tau} = \frac{1}{g_\tau^2}\,.
\label{eq_beta_tau}
\end{align}

In the second term in the action~\eqref{eq_S}, the series of the double derivative~\eqref{eq_d_phi_plaquette} appears to be in the higher order in the lattice spacing: 
\begin{align}
    \phi_{{\boldsymbol{x}},xy} =  a^{5/2} \partial_x \partial_y \phi({\boldsymbol{x}}) + O(a^{9/2})\,.
\label{eq_phi_scaling_2}
\end{align}
Therefore, the second term will scale in the continuum limit as $O(a^5)$, which has a higher power in $a$ as required by the volume scaling $a^3$ for a unit lattice cube. Thus, the second term in the action~\eqref{eq_S} should naively vanish in the continuum limit for smooth lattice configurations. There are, however, three arguments against this conclusion, which allow us to consider the model~\eqref{eq_S} seriously. 

Firstly, one can consider the model~\eqref{eq_S} as a pure lattice model which has its experimentally interesting counterparts in the condensed matter application. In this case, the lattice spacing $a$ is determined by an interatomic distance of the underlying crystal lattice, which provides a natural ultraviolet cutoff.

Secondly, in the presence of vortices, the derivatives become singular, and the series of the kind~\eqref{eq_phi_scaling_1} and \eqref{eq_phi_scaling_2} ---valid for smooth field configurations--- are no more valid. A related example can be given by compact electrodynamics~\cite{Polyakov:1976fu}, which contains singular configurations of the gauge field called Abelian monopoles. While a naive continuum limit, achieved with expansions similar to Eqs.~\eqref{eq_phi_scaling_1} and \eqref{eq_phi_scaling_2}, is trivial, the model itself is far from being simple. It features confinement and mass gap generation phenomena~\cite{Polyakov:1976fu}, a non-trivial phase structure~\cite{Coddington:1986jk,Chernodub:2001ws}, and the non-perturbative Casimir effects~\cite{Chernodub:2017mhi}.

Thirdly, similarly to gauge theories in three spacetime dimensions, a nontrivial continuum action even for smooth configurations of $\phi({\boldsymbol{x}})$ can also be achieved by the following rescaling of the coupling constant:
\begin{align}    
    \beta_{xy} = \frac{1}{g_{xy}^2 a^2}\,.
\label{eq_beta_xy}
\end{align}
This type of rescaling is also typical for the mentioned gauge theories in three spacetime dimensions. As a result, the spatial dynamics of the field $\phi$ are controlled by the parameter $g_{xy}$, which has the dimension of mass. One arrives at the second term in the action~\eqref{eq_S}:
\begin{align}
    \beta_{xy} & \sum_{\boldsymbol{x}} [1 - \cos (\Delta_x \Delta_y \phi_{\boldsymbol{x}})] \nonumber\\
    & \to \frac{1}{2 g^2_{xy}} \int d^3 x \, \bigl[\partial_x \partial_y \phi({\boldsymbol{x}})\bigr]^2\,,
    \label{eq_continuum_2}
\end{align}
\corr{Similarly to the comment made after Eq.~\eqref{eq_continuum_1}, we stress that the configutation is assumed to be smooth thus excluding the presence of vortices.}

Combining now \corr{the continuum actions given for smooth, vortex-free configurations,} Eqs.~\eqref{eq_continuum_1} and \eqref{eq_continuum_2}, we get for the lattice action~\eqref{eq_S} the following continuum counterpart:
\begin{align}
S_\phi = \int d^3 x \biggl[
\frac{1}{2 g^2_\tau}  \bigl(\partial_\tau \phi({\boldsymbol{x}})\bigr)^2
+ \frac{1}{2 g^2_{xy}} \bigl(\partial_x \partial_y \phi({\boldsymbol{x}})\bigr)^2\biggr]\,,
\label{eq_L_continuum_naive}
\end{align}
which represents an analogue of the action~\eqref{eq_L_continuum_rescaled} discussed earlier. Let us here stress the word "naive" in relation to action~\eqref{eq_L_continuum_naive}, which is put to imply that vortices are \corr{intentionally disregarded} in this formulation. In our work, we assume that the vortices are present, implying that the field $\phi$ in the naive formulation~\eqref{eq_L_continuum_naive} is a compact scalar field that should be considered as a (necessarily compact) phase of a more fundamental field $U = e^{i \phi}$, rather than an independent field itself. After all, it is the winding of the compact phase in the bosonic condensate $\Phi$ that leads to the formulation of vortices in liquid helium~\cite{volovik2003universe, Kleinert2008}. 

The compactness of the model and the existence of the vortices appear in the singularities of the field $\phi$, which implies, mathematically, that the derivatives do not commute~\cite{Kleinert2008}:
\begin{align}
    [\partial_\mu, \partial_\nu] \phi({\boldsymbol{x}}) =  \varepsilon_{\mu\nu\alpha} 
    v^\mu({\boldsymbol{x}})\,,
    \label{eq_dd}
\end{align}
where $\varepsilon_{\mu\nu\alpha}$ is the totally antisymmetric Levi-Civita tensor (which highlights a special role played by the dual lattice in the lattice construction). In Eq.~\eqref{eq_dd}, which is a continuum analogue of the lattice construction~\eqref{eq_v_P}, the one-dimensional worldline of a vortex 
\begin{align}
    v^\mu({\boldsymbol{x}}) = \frac{1}{2\pi} \int d \xi \frac{\partial {\bar x}^\alpha(\xi)}{\partial \xi} \delta^{(3)} \bigl( \boldsymbol{x} - {\bar {\boldsymbol{x}}}(\xi)\bigr)\,,
\end{align}
is parameterized by the function $\bar{\boldsymbol{x}} = \bar{\boldsymbol{x}}(\xi)$ of the continuous parameter~$\xi$. We remind that in three spacetime dimensions, the vortices are particle-like objects. Their conservation is supported by the identity $\partial_\mu v^{\mu} (\boldsymbol{x}) = 0$ which is analogue of the lattice identity~\eqref{eq_dual_conservation}.

The difference between compact and non-compact formulations of the same theory can be readily understood in well-studied quantum electrodynamics, which in both cases can be formally described by the same (Euclidean) Lagrangian ${\mathcal L}_{A} = (1/4) F_{\mu\nu}^2$. \corr{In the non-compact case, the model is a featureless Maxwell theory which possesses only one gapless phase. In the compact formulation of the mode, the field configurations contain monopole like singularities, which lead to mass gap generation and confinement of charges~\cite{Polyakov:1976fu}. In particular, the phase structure of the compact model becomes rather nontrivial implying the existence of two phases: the gapless (deconfinement) phase at high temperatures as well as at weak coupling and the gapped (confining) phase at low temperatures as well as at strong coupling~\cite{Coddington:1986jk,Chernodub:2001ws}. The aim of this paper is, in particular, to elucidate the phase structure of the compact version of the scalar model~\eqref{eq_L_continuum_naive}, where the singular (vortex) configurations are allowed similarly to the standard XY model.} 

Coming back to the field redefinition in the vortex-less limit, the field $\phi$ and the coordinates ${\boldsymbol{x}}$ can be formulated terms of their dimensionless counterparts, $\phi_0$ and ${\boldsymbol{x}}_0$, respectively:
\begin{align}
\phi = \frac{\phi_0}{\sqrt{g_\tau g_{xy}}}\,,
\quad
\tau = \frac{\tau_0}{g_\tau g_{xy}}, 
\quad\
(x,y) = \frac{(x_0,y_0)}{g_{xy}}, 
\label{eq_rescaling}
\end{align}
in terms of which the non-compact (``n.c.'') action~\eqref{eq_L_continuum_naive} can be cast into a coupling-less form~\eqref{eq_L_continuum_rescaled} mentioned already in the Introduction.\footnote{In Eq.~\eqref{eq_L_continuum_rescaled}, we remove the subscripts ``0'' from the dimensionless rescaled variables defined in Eq.~\eqref{eq_rescaling}.} At zero temperature, the phase diagram of this non-interacting model consists only of one point and is, therefore, trivial.\footnote{A non-nontrivial counter-example described by a coupling-less Lagrangian is represented by Yang-Mills theory in three spacetime dimensions. In this theory, the gauge coupling is a dimensionful quantity that can be removed by rescaling. However, contrary to the non-interacting model~\eqref{eq_L_continuum_rescaled}, Yang-Mills theory possesses highly nontrivial phenomena such as confinement of color and mass gap generation~\cite{Teper:1998te}. Emerging from non-perturbative gluon interactions, the mechanism of these effects is not well understood despite certain advances in understanding the mass scales of the theory~\cite{Karabali:1995ps, Karabali:1996je, Karabali:1998yq, Chernodub:2018pmt, Karabali:2018ael}.} 

Notice that the above statements apply only to the non-compact form of the model, which does not maintain vortex topological defects. On the contrary, the compact form of the action possesses vortex excitations. The vortex density sets another dimensional scale in this model, thus making it impossible to apply the rescaling~\eqref{eq_rescaling}. Therefore, the compact can possess a nontrivial phase diagram, which is one of the focus points of this paper.

\subsection{Numerical methods}

In our paper, we address the statistical properties of the model~\eqref{eq_S} using numerical simulations based on the Monte Carlo techniques. We simulate the lattice action~\eqref{eq_S} at the cubic lattices $L^3$ with various extensions $L = 8, \dots, 64$, and periodic boundary conditions in all directions. For simulations, we used the Hybrid Monte Carlo algorithm and applied Nvidia CUDA API to provide calculations using GPU. We collected statistics of 100 configurations per each point of a heatmap (25600 points in total per each heatmap). The results of our numerical simulations are described in the rest of the paper.

\begin{figure*}[!htb]
\centering
  \includegraphics[width=0.99\linewidth]{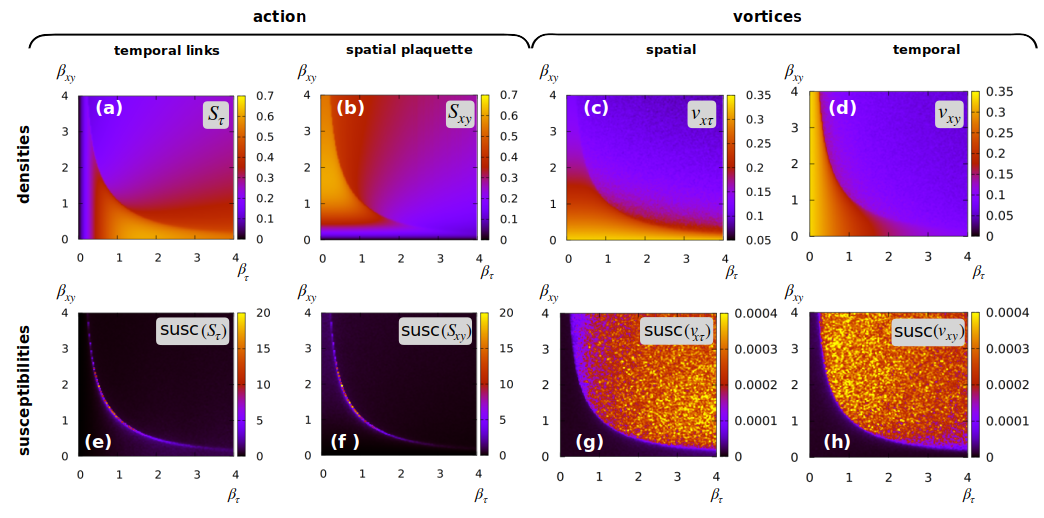}
  \caption{The expectation values of the observables (the upper row) and their susceptibilities (the lower row) associated with the action densities (the left side) and the vortex densities (the right side). The upper row of the observables includes the action terms for (a) temporal links $\avr{S_\tau}$, Eq.~\eqref{eq_S_tau} and (b) spatial plaquettes $\avr{S_{xy}}$, Eq.~\eqref{eq_S_xy}; the densities~\eqref{eq_v_P} of (c) the spatial vortex trajectories, $v_x$ and $v_y$, piercing the spatial-temporal plaquettes $\avr{|v_{x\tau}|}$ and (d) temporal vortices $v_\tau$ that pierce purely spatial plaquettes $\avr{|v_{xy}|}$. The susceptibilities~\eqref{eq_susc_O}, ${\rm susc}({\cal O}) = \avr{{\cal O}^2} - \avr{{\cal O}}^2$, are shown in the lower row in the plots (e)-(h) below the plots (a)-(d) of the expectation values of the corresponding densities $\avr{\cal O}$.}
  \label{fig_action_vortex}
\end{figure*}

\section{Phase diagram}
\label{sec_results_phase}

\subsection{The critical phase transition line}

\subsubsection{Action density and its susceptibility}

A first suggestive hint on the phase structure of a model can be obtained from a simple analysis of the expectation values of separate terms that enter the action of the model. In our case, the lattice action~\eqref{eq_S} is a sum~\eqref{eq_S_phi} of two qualitatively different contributions: 
\begin{itemize}
    \item the link term $S_\tau$ in Eq.~\eqref{eq_S_tau}, which describes the standard coupling of the fields belonging to different time slices and 
    \item the plaquette term $S_{xy}$, Eq.~\eqref{eq_S_xy}, which describes the interaction of the field within each time slice. 
\end{itemize}
The volume-normalized expectation values of these contributions are shown, respectively, in Figs.~\ref{fig_action_vortex}(a) and \ref{fig_action_vortex}(b) in the plane of the couplings $(\beta_\tau,\beta_{xy})$. 

Different parts of the action appear to dominate different regions of the coupling space $(\beta_\tau,\beta_{xy})$. The general dependence of the action term on the couplings has a rather complex and complementary form: the link variable $\avr{S_\tau}$ takes its maximum at lower $\beta_{xy}$ and larger $\beta_\tau$, while the plaquette term $\avr{S_{xy}}$ is enhanced at the different corner of the coupling space at lower $\beta_{\tau}$ and larger $\beta_{xy}$. Both action densities have signatures of (phase) transitions, which are clearly visible in both plots. 

\corr{A nontrivial phase structure of the Villain representation of the plaquette model has been observed numerically in Ref.~\cite{Paramekanti2002}.} \corr{We also notice the important difference between the lattice formulation of the Wilson form that we use in our paper~\eqref{eq_S} and the modified Villain formulation which has been used in Ref.~\cite{Gorantla:2021svj}. While our model admits the presence of the vortex excitations, the one of Ref.~\cite{Gorantla:2021svj} does not. These models are only equivalent to each other in a certain region of the parameter space, where vortices are suppressed. Moreover, the modified Villain model of Ref.~\cite{Gorantla:2021svj} features a trivial phase diagram as it always appear to reside gapless phase contrary to the model studied in our paper. This crucial different appears as a result of the vortex dynamics as we discuss below.}

In general, the position(s) of the (phase) transition(s) in a statistical field theory can be probed with the help of the susceptibility
\begin{align}
    {\rm susc}({\cal O}) = \avr{{\cal O}^2} - \avr{{\cal O}}^2\,,
    \label{eq_susc_O}
\end{align}
of a bulk quantity ${\cal O} = {\cal O}(\phi)$ which is sensitive to a particular transition. At a (phase) transition, the statistical fluctuations of the fields $\phi$ are usually the largest compared to the regions of the model couplings, which are located farther from the transition point. The scale of fluctuations of a quantity $\cal O$ is captured by its susceptibility~\eqref{eq_susc_O}, implying that the transition points should be associated with the maxima of \eqref{eq_susc_O}, when computed, necessarily, in a finite volume. In order to better resolve the position of the phase transition, the quantity $\cal O$ should be chosen in such a way that it catches the relevant critical fluctuations of the fields. For example, in the case of a phase transition driven by a (spontaneously) broken symmetry, the quantity $\cal O$ should preferably be an order parameter associated with the corresponding symmetry. However, on the practical side, as the field fluctuations are enhanced at the transition point, almost any thermodynamic quantity can be used to probe the location of the transition point via the maximum in its susceptibility. 

We show the susceptibilities~\eqref{eq_susc_O} of the temporal link action term $\avr{S_\tau}$ and the spatial plaquette action term $\avr{S_{xy}}$ in Figs.~\ref{fig_action_vortex}(e) and \ref{fig_action_vortex}(f), respectively, below the plots of their corresponding expectation values in Figs.~\ref{fig_action_vortex}(a) and \ref{fig_action_vortex}(b). The link and plaquette action susceptibilities have similar features that undoubtedly point to the existence of a single phase transition that has already been spotted in the expectation values of the lattice actions in Figs.~\ref{fig_action_vortex}(a) and \ref{fig_action_vortex}(b): the phase transition extends from a low-$\beta_\tau$ region at high $\beta_{xy}$ to a low-$\beta_{xy}$ values at high $\beta_{\tau}$.  

\begin{figure*}[!htb]
\centering
  \includegraphics[width=0.95\linewidth]{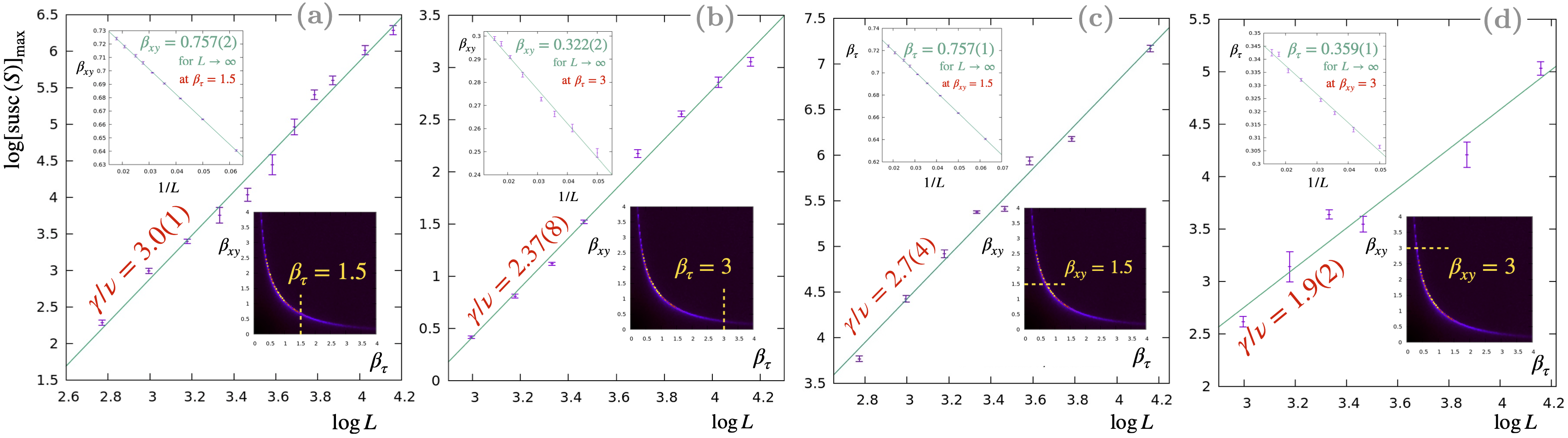}
  \caption{The maximum of the susceptibility~\eqref{eq_susc_O} of the action density~\eqref{eq_S}, ${\cal O} = S/L^3$, as the function of the lattice length $L$ is shown in the log-log scale. The lines represent the best fits for the scaling behaviour~\eqref{eq_chi_max} with the critical exponent $\gamma/\nu$ shown in the plots. The maximum of the susceptibility is taken along the cut lines in the $(\beta_{\tau},\beta_{xy})$ coupling plane that crosses the phase transition as shown in the bottom-right inset of each plot. 
  The left-top insets show the best fit~\eqref{eq_beta_L_fit} of the corresponding pseudocritical coupling constant vs the inverse size of the system, $1/L$. The critical values of the coupling constants, shown in the insets, are summarized in Eq.~\eqref{eq_pairs_beta_c}.}
  \label{fig_susceptibility}
\end{figure*}

It turns out that the peaks of the susceptibilities of the link and plaquettes actions, shown in Figs.~\ref{fig_action_vortex}(a) and \ref{fig_action_vortex}(b), overlap with each other. This implies the existence of a single {}phase transition that separates two different phases in the model. Before clarifying the nature of these two phases, we study the order of the phase transition and the functional dependence of the critical line that separates them.

\subsubsection{Order of the phase transition}

What is the order of the phase transition that separates the two phases? To this end, we analyzed the susceptibility~\eqref{eq_susc_O} of the action density $S/L^3$ of the model~\eqref{eq_S} at four points along the line of the phase transition. Namely, we calculated this quantity along the two vertical lines given by the fixed temporal link couplings $\beta_\tau = 1.5$ and $\beta_\tau = 3.$ with varying spatial plaquette coupling $\beta_{xy}$ and another two horizontal lines given by the fixed vector link couplings $\beta_{xy} = 1.5$ and $\beta_{xy} = 3.$ with varying spatial plaquette coupling $\beta_{\tau}$. These lines are shown in the right-bottom insets of Figs.~\ref{fig_susceptibility}(a)-(d).

The maxima of these susceptibilities are then determined by fitting their peaks by a Gaussian function of the corresponding $\beta$ coupling (not shown), with the central value providing us with the behavior of the pseudo-critical transition coupling $\beta_{xy,c} = \beta_{xy,c}(L)$ and $\beta_{\tau,c} = \beta_{\tau,c}(L)$, respectively. It turns out that all the pseudo-critical values of the coupling constants can be fitted by an anticipated inverse-linear behavior,
\begin{align}
    \beta_{a,c}(L) = \beta^{(\infty)}_{a,c} + \frac{C_a}{L}\,,\qquad a =\tau, xy\,,
    \label{eq_beta_L_fit}
\end{align}
valid for $L \geqslant 8$. In the function above, the fit parameters are the asymptotic thermodynamic value 
of the coupling constant, $\beta^{(\infty)}_{a,c}  = \lim_{L \to \infty} \beta_{a,c}(L)$, and an inessential prefactor $C_a$. The best fits~\eqref{eq_beta_L_fit} are shown in the left-top insets of Figs.~\ref{fig_susceptibility}(a)-(d). The pairs of the thermodynamic ($L \to \infty$) critical values are as follows:
\begin{align}
    \bigl(\beta_{\tau_c}^{(\infty)}, \beta_{xy,c}^{(\infty)}\bigr)
  = & \bigl[1.5, 0.757(2)\bigr]^{{\rm (1st)}}, \quad \bigl[3, 0.322(2)\bigr]^{{\rm (cnt)}}, \nonumber \\ 
    & \bigl[ 0.757(1) , 1.5 \bigr]^{{\rm (cnt)}}, \quad \bigl[0.359(1), 3 \bigr]^{{\rm (cnt)}}. 
    \label{eq_pairs_beta_c}
\end{align}
The order of the phase transition for the determined critical coupling pairs is indicated in Eq.~\eqref{eq_pairs_beta_c} as the superscripts ``1st'' or ``cnt''. The transition order is determined as follows. 

The main plots in Figs.~\ref{fig_susceptibility}(a)-(d) show the maximum susceptibility of the action density~\eqref{eq_S} as the function of the lattice length $L$. The way how the peak of the susceptibility $\chi_{\rm max}$ diverges as the system approaches the thermodynamic limit, $L \rightarrow \infty$, reveals whether the phase transition is first-order or continuous (second-order, higher or of the BKT type~\cite{Berezinsky:1970fr, Berezinsky:1972rfj, Kosterlitz1973}). According to the finite-size scaling analysis, valid at large $L$,
\begin{align}
    \chi_{\rm max} \propto L^{\gamma/\nu}\,,
\label{eq_chi_max}
\end{align}
where we adopted the traditional notation suitable for the second-order phase transitions: $\gamma$ is the susceptibility critical exponent, and $\nu$ is the correlation length exponent. All four plots in  Fig.~\ref{fig_susceptibility} indicate that our data are perfectly described by the scaling behavior~\eqref{eq_chi_max}.

A first-order phase transition is characterized by a thermodynamic discontinuity of various bulk observables (in our case, it is the action density) and finite values of all correlation lengths at the transition point. In this case, the susceptibility peak scales with the volume of the statistical system, $\chi_{\rm max} \propto L^3$, implying $\gamma/\nu = 3$ in Eq.~\eqref{eq_chi_max}. This behavior is observed for the first pair of the critical couplings in Eq.~\eqref{eq_pairs_beta_c} shown in Fig.~\ref{fig_susceptibility}(a). 

For a continuous phase transition, the critical exponent in the scaling behavior~\eqref{eq_chi_max} is a quantity that is smaller than the dimensionality of the system, $\gamma/\nu < 3$. This is the case for the remaining three pairs of the critical couplings in Eq.~\eqref{eq_pairs_beta_c} shown in Figs.~\ref{fig_susceptibility}(b)-(d).

\begin{figure*}[t]
\centering
  \includegraphics[width=0.99\linewidth]{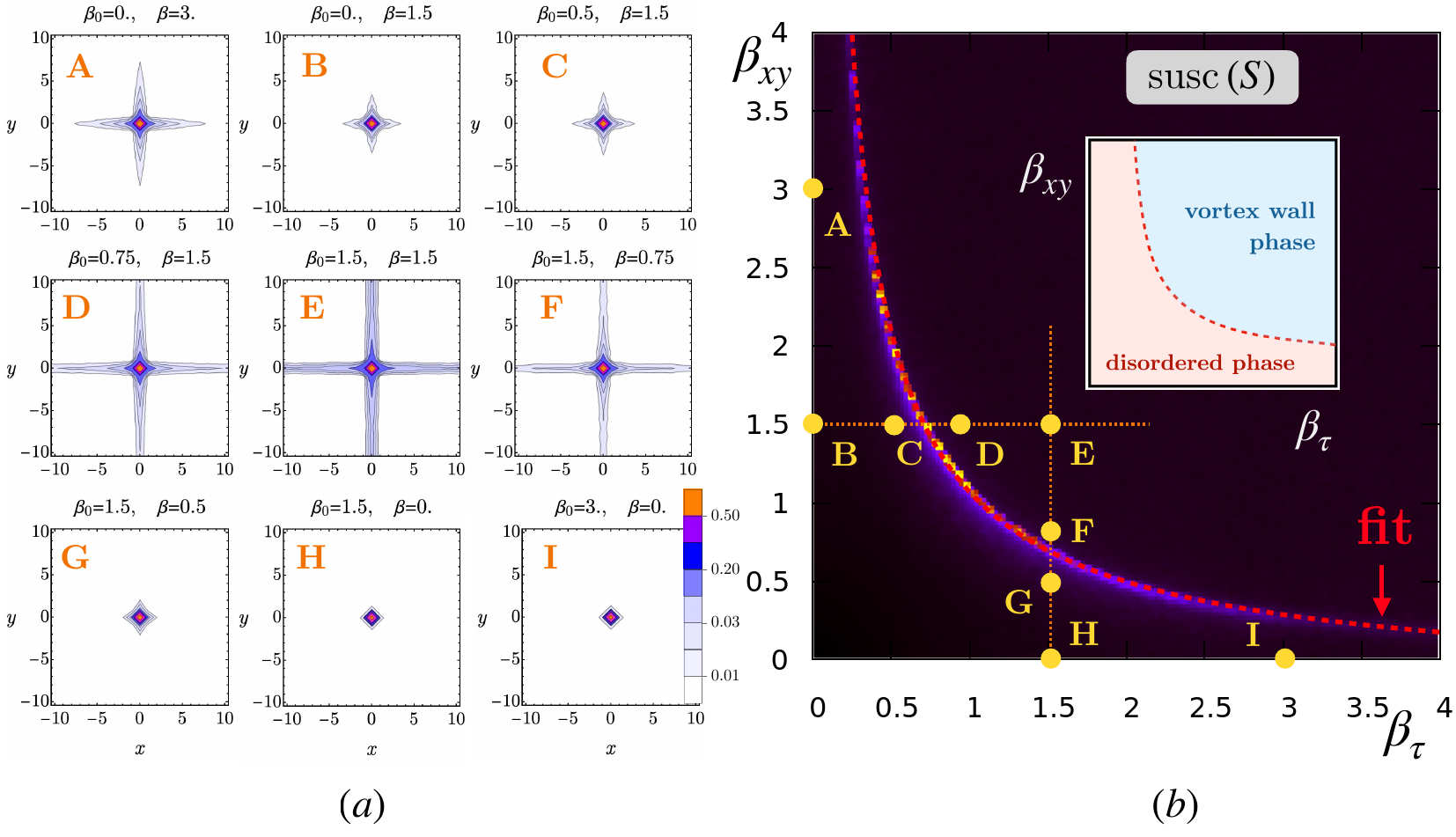 }
  \caption{(a) The correlation function~\eqref{eq_corr_fun}, \eqref{eq_corr_fun_mean} of the temporal vortex trajectories is shown in the ($x$,$y$) plane for nine parameter sets A-I of the lattice couplings $\beta_\tau$ and $\beta_{xy}$. The positions of each of the sets are shown at the $(\beta_\tau, \beta_{xy})$ phase diagram at the right panel (b), which also shows the density plot of the susceptibility of the full lattice action density~\eqref{eq_S}. The best fit of the line represented by the maximal value of the susceptibility of the action is given by the phenomenological formula~\eqref{eq_fit} with the best-fit parameters~\eqref{eq_best_fits}. The inset in (b) highlights the nature of the phases shown in the main figure.} 
  \label{fig_correlations}
\end{figure*}

\corr{Notice that the connection with the renormalization group, as appears from the suggestive finite-scaling behaviour~\eqref{eq_chi_max}, is not entirely straightforward. Fractonic theories are known to exhibit a nontrivial mixing between ultraviolet (UV) and infrared (IR) degrees of freedom, which significantly complicates the application of standard renormalization group techniques~\cite{Lake2022}. }

\subsubsection{The critical line}

Our numerical results show that the phase transition line can be represented with excellent accuracy by the following phenomenological dependence:
\begin{align}
    \beta_{\tau,c}(\beta_{xy}) = \frac{A_1}{\beta_{xy}^{b}} + \frac{A_2}{\beta_{xy}^{2 b}}
    \qquad \text{[phase transition]}\,,
    \label{eq_fit}
\end{align}
where the prefactors $A_1$ and $A_2$, and the exponent $b$ are defined with the help of a fitting procedure. The best fit of the susceptibility~\eqref{eq_susc_O} of the total action density, ${\cal O} = S/L^{3}$, Eq.~\eqref{eq_S_phi}, by the functional dependence~\eqref{eq_fit} is shown in Fig.~\ref{fig_correlations}(right). The best-fit curve is determined by the following best-fit parameters:
\begin{align}
    A_1 & = - 0.80(6)\,, \nonumber \\
    A_2 & = \phantom{-} 1.86(36)\,,\ \qquad \text{[at $32^3$ lattice]} 
    \label{eq_best_fits}\\
    b & = \phantom{-} 0.362(6)\,. \nonumber 
\end{align}

The functional form of the critical curve~\eqref{eq_fit} and the numerical values of the best-fit parameters~\eqref{eq_best_fits} suggest that at small values of the plaquette coupling $\beta_{xy}$, the critical curve slowly approaches the vertical $\beta_{xy} = 0$ axis: $\beta_{\tau,c}(\beta_{xy}) \simeq 1.87 / \beta_{xy}^{0.72}$ as $\beta_{xy} \to 0$. On the other hand, the negative value of the coefficient $A_1$ in Eq.~\eqref{eq_best_fits} indicates that the critical curve $\beta_{\tau,c} = \beta_{\tau,c}(\beta_{xy})$ hits the axis $\beta_\tau = 0$ at a large value of the coupling constant, $\beta^{\rm E}_{xy} = (-A_2/A_1)^{1/b} \sim 10$, thus implicitly suggesting the presence of the endpoint $(\beta_\tau, \beta_{xy})^{\rm E} = (0, \beta^{\rm E}_{xy})$ at the lattice $32^3$. The position and the very existence of this transition endpoint may be a finite-volume feature supported by the two-term truncation of the phenomenological dependence~\eqref{eq_fit}. 

\corr{One should notice that in the limit of the strong time-like coupling, $\beta_\tau \to 0$, the 2+1-dimensional model~\eqref{eq_S} reduces to a stack of $L$ disconnected classical XY 1+1-dimensional models. The Mermin–Wagner theorem~\cite{Mermin:1966fe} forbids spontaneous breaking of a continuous symmetry in one spatial dimension. In this limit, the system is characterized  by a quasi-long-range order in the sense that any perturbation can destroy it, but the ground state itself is ordered. This property implies that, in the strict thermodynamic limit, no (phase) transitions at $\beta_\tau = 0$ axis of the phase diagram should take place. At finite temperature, however, one could expect a crossover transition from disordered (finite $T$) to ordered (at $T=0$) behavior which highlights the fragility of the system against thermal fluctuations that lead to exponential decay of correlations~\cite{Hohenberg1967}.}

\corr{On the other hand, one spatial length $L$ in the 2+1-dimensional model in the volume $L^3$ can be associated, in the $\beta_\tau \to 0$ limit, with an imaginary time dimension of the 1+1 XY model. Therefore, small thermal fluctuations could be present in the resulting 1+1 dimensional finite-volume system which could justify the existence of crossover between long-range order for all $T >0$ and the quasi-long-range order at $T=0$. In short, one could expect that the existence of the transition endpoint at $\beta_\tau \to 0$ is the result of this finite-volume (finite-temperature) crossover.}

\subsection{Bulk vortex properties: a signature of an "over-disordered" phase}

\subsubsection{Vortex density}

Due to the difference in the nature of spatial and temporal interactions of scalar fields $\phi_{\boldsymbol x}$ highlighted by the form of the lattice action~\eqref{eq_S}, one can suggest that spatial $xy$ and, separately, spatial-temporal $x\tau$ and $y\tau$ vortex plaquettes could carry largely uncorrelated information about the properties of the model. As we discussed above, the spatial plaquette $xy$ is pierced by the vortex line parallel to the imaginary time axis $\tau$. Therefore, the spatial plaquette $xy$ corresponds to a timeline vortex trajectory. Similarly, the spatial-temporal $x\tau$ and $y\tau$ vortex plaquettes describe purely spatial segments of the vortex lines that are oriented along the $y$ and $x$ axes, respectively.

Technically, we construct both temporal and spatial vortex lines using the same prescription given in Eqs.~\eqref{eq_bar_phi} and \eqref{eq_v_P}. The densities of the spatial and temporal vortex segments are defined as follows:
\begin{align}
    \text{spatial:} \label{eq_xtau}\\[-3mm]
    v_{i\tau} \equiv & \avr{|v_{i\tau}|} = \frac{1}{2 L^3} \avr{\sum_{\boldsymbol{x}} \left| v_{\boldsymbol{x},i\tau} \right|}\,, \qquad i = x,y \,, \nonumber \\
    \text{temporal:} \nonumber \\[-3mm]
v_{xy} \equiv & \avr{|v_{xy}|} = \frac{1}{L^3} \avr{\sum_{\boldsymbol{x}} \left| v_{\boldsymbol{x},xy} \right|}\,, 
\label{eq_xy}
\end{align}
(notice that $\avr{|v_{x\tau}|} = \avr{|v_{y\tau}|}$ at large statistics). The behavior of the spatial~\eqref{eq_xtau} and temporal~\eqref{eq_xy} vortex densities in the coupling plane is shown in Figs.~\ref{fig_action_vortex}(c) and \ref{fig_action_vortex}(d), respectively. 

The properties of both vortex densities point to the existence of two phases that we have already noticed in studying the properties of the action densities. At the large coupling domain, with $\beta_\tau \sim \beta_{xy}$ both taking large values, the vortex density gets relatively low. While --based on the
analogy with the XY model-- a small vortex density is undoubtedly expected to occur in the region with large $\beta$’s, it is still impressive to notice that the vortex density takes a relatively moderate value, with $\avr{|v_{\tau}|} \sim \avr{|v_{xy}|} \sim 0.1$. This property points to an exotic nature of the large-$\beta$'s phase, which we will discuss slightly later. 

Figure~\ref{fig_action_vortex}(c) indicates that if the temporal (link) lattice coupling $\beta_\tau$ gets small, then the spatial vortex density suddenly increases. This behavior can also be readily understood from the properties of the $XY$ model: a decrease in the temporal lattice coupling leads to more fluctuations/disorder of site variables along the temporal links, which supports the spatial proliferation of the vortex trajectory. The sudden change in the vortex properties is undoubtedly associated with the (phase) transition, as it was already noticed in the behavior of the action densities. Analogously, if the spatial (plaquette) lattice coupling $\beta_{xy}$ becomes smaller than the critical value, the density of temporal vortex segments gets increased, as is clearly seen in Fig.~\ref{fig_action_vortex}(d). 

The densities of spatial and temporal segments of the vortex trajectories, shown in Figs.~\ref{fig_action_vortex}(c) and \ref{fig_action_vortex}(d), visually correlate with the densities of temporal and spatial parts of the actions shown in Figs.~\ref{fig_action_vortex}(a) and \ref{fig_action_vortex}(b). The data on vortex properties also suggests (if the results are treated straightforwardly) the existence of disordered [at the left side of the critical line~\eqref{eq_fit}, $\beta_{\tau} < \beta_{\tau,c}$] and more ordered [at the right side of the critical line~\eqref{eq_fit}, $\beta_{\tau} > \beta_{\tau,c}$] phases separated by a transition. However, the most transparent manifestation of a phase transition line is provided not by the densities of the actions but rather by their susceptibilities, as shown in Figs.~\ref{fig_action_vortex}(e) and \ref{fig_action_vortex}(f). The line where the susceptibilities of both parts of the action take their maxima indicates the onset of a phase transition in the system. What happens with the vortex susceptibilities (the susceptibilities of the vortex densities)?

\subsubsection{Susceptibility of the vortex density}

The susceptibilities of the spatial and temporal vortex trajectories, shown in Figs.~\ref{fig_action_vortex}(g) and \ref{fig_action_vortex}(h), respectively, offer us a surprise. Indeed, lattice simulations of the XY model and similar models possessing vortex defects consistently show a pronounced peak in vortex susceptibility at a phase transition. This peak is one of the critical signatures of the transition and is often used in numerical studies to determine the transition temperature accurately (see, for example, Ref.~\cite{Bittner2005}). However, the vortices in the plaquette-XY model~\eqref{eq_S} show a completely different behavior. Instead of exhibiting a single peak line similar to the susceptibilities of the parts of the action, presented already in Figs.~\ref{fig_action_vortex}(e) and \ref{fig_action_vortex}(f), both vortex susceptibilities almost vanish at the phase located at the left side of the critical line~\eqref{eq_fit}, $\beta_{\tau} < \beta_{\tau,c}$, and take its almost flat maximum at the right side of the transition line. 

Moreover, at the right-side phase, $\beta_{\tau} > \beta_{\tau,c}$, the susceptibility does not exhibit a smooth behavior, showing a certain degree of disorder easily visible in both plots of Figs.~\ref{fig_action_vortex}(e) and \ref{fig_action_vortex}(f). We have found that a many-fold increase in the statistics of simulation does not entirely remove nor substantially reduce these visible fluctuations, which seem to be a genuine part of the system. 

The absence of a peak in the susceptibility at the phase transition may be associated with the particular dynamics of semiclassical fluctuations that exist in one phase and are suppressed in the other one. The fluctuations of the fields may be of a topological nature, generated by monopoles or vortices in field theories. The dynamics of these topological defects create a disorder in the fields at the phase where the topological objects are present. A good example is given by electroweak vortices in the vacuum of the Standard Model of particle physics, where the susceptibility of a scalar field does not demonstrate a single localized peak across a phase transition. On the contrary to the expectations, it stays essentially constant in a vortex-dominated phase~\cite{Chernodub:2022ywg}, exhibiting behavior similar to the picture shown in Figs.~\ref{fig_action_vortex}(e) and \ref{fig_action_vortex}(f). 

In the XY-type models, the susceptibility of the vortex density has a peak at the transition point, implying that the susceptibility tends to decrease and becomes a small quantity as we move further from the transition point, either at the one size or at the other one. For example, in the two-dimensional XY model, the vortex density is small at the low-temperature ordered phase, which implies the suppression of the vortex susceptibility. While the vortex density is high at the high-temperature disordered phase, the vortex density exhibits random small fluctuations over the dense vortex state and, correspondingly, implies a low value of the vortex susceptibility likewise. A significant value of the vortex susceptibility in the whole $\beta_\tau > \beta_{\tau,c}$ phase may indicate that the fluctuations of the densities of the vortex trajectories are not random in this phase. We discuss the properties of the vortices in the plaquette-XY model in more detail in the next section.

\section{Vortices}
\label{sec_results_vortex}

\subsection{Vortex correlation function}

The statistical properties of the vortices can also be analyzed with the help of the correlation functions of their trajectories. In our paper, we study the correlation function of the temporal segments of the vortex worldlines:
\begin{align}
    C_v({\boldsymbol{x}}) = \frac{1}{L^3}\sum_{{\boldsymbol{y}}} \Bigl[\avr{|v_\tau(\boldsymbol{x} + \boldsymbol{y})| |v_\tau(\boldsymbol{y})|} - 
    \avr{|v_\tau(\boldsymbol{y})|}^2 \Bigr]\,,
    \label{eq_corr_fun}
\end{align}
where the two-point correlation function is averaged over the whole lattice volume $L^3$. The correlator is normalized in such a way that, 
\begin{align}
    \lim_{|\boldsymbol{x}| \to \infty} C_v({\boldsymbol{x}}) = 0\,,
    \label{eq_C_vanishing}
\end{align}
for an asymptotically isotropic, at large distances, phase such as the disordered, high-temperature phase of the XY model. On a practical level, we also perform, in order to improve the measurement statistics, the averaging of the correlation function over the imaginary time~$\tau$:
\begin{align}
    C_v(x,y) = \sum_{\tau = 0}^{L-1} C_v(x,y,\tau)\,.
    \label{eq_corr_fun_mean}
\end{align}
Notice that an equal-time correlation function, $C_v(x,y, 0)$, has a very similar {\it qualitative} behavior as the time-averaged function~\eqref{eq_corr_fun_mean}. Other two-point functions, including spatial-temporal and spatial-spatial correlations, are also identical to the correlation between the temporal vortex densities~\eqref{eq_corr_fun}, so they are not shown in our paper. The two-dimensional density plots of the correlation function~\eqref{eq_corr_fun} and \eqref{eq_corr_fun_mean} are shown in Fig.~\ref{fig_correlations}(a). 

The correlation function~\eqref{eq_corr_fun} in the disordered phase at $\beta_\tau < \beta_{\tau,c}$ exhibits an expected decline with the increase in its arguments $x$ or $y$. However, at points A, B, and C, the function $C_v(x,y)$ reveals a striking breaking of the rotational symmetry group in the $xy$ plane. This feature is a rather unexpected property of the plaquette-XY model because the usual XY model, like many ordinary and physically relevant models, exhibits restoration of the rotational symmetry in the disordered phase characterized by a high density of topological defects. Therefore, in a disordered and vortex-dominated phase, we would expect an isotropy of correlation functions at large distances. 

The isotropy of the vortex correlation functions could also be expected if one invokes a close resemblance of the vortex dynamics with the roughening property of the boundaries (domain walls) in spin models. For example, In the 2D Ising model, at temperatures below the critical temperature, interfaces between spin-up and spin-down regions are smooth and break rotational symmetry. The correlation functions associated with the boundaries in this phase are anisotropic. However, at high temperatures (above the roughening transition in the 3D Ising model), the interfaces become much less disordered, and the correlation functions become isotropic, reflecting the restored rotational symmetry. The roughening transition marks the change from a broken rotational symmetry in the smooth phase to a restored rotational symmetry in the rough phase.~\cite{Weeks1973}

The source of the anisotropy of the vortex correlators in the plaquette-XY model can be traced back to the presence of the plaquette term that breaks the rotational symmetry in the $xy$ plane at the level of the fields. This statement is evident from the comparison of the lattice action~\eqref{eq_S} with its continuum counterpart~\eqref{eq_L_continuum_rescaled}. Consistently, the symmetry-breaking plaquette term plays a pronounceable role in the action at large values of the plaquette coupling $\beta_{xy}$, which correspond precisely to points A, B, and C. On the contrary, a decrease in the coupling $\beta_{xy}$ diminishes the anisotropy in the correlation function ({\it cf.} correlations in A {\it vs.} the ones in B), while the increase in the link coupling $\beta_\tau$, provided the systems remain in the disordered phase, does not reveal a significant effect on the anisotropy (points B vs. C). Expectedly, at small values of $\beta_{xy}$ (points G, H, and I), the anisotropy stays minimal in the disordered phase. 

The qualitative properties of the vortex correlations, however, get entirely changed in the other phase at $\beta_\tau > \beta_{\tau,c}$, where the correlation function has a cross-like long-ranged pattern. As it is seen at the points D, E, and F, the correlation function reaches non-zero plateaus at two axes given by $x \to 0$ and large $|y|$ as well as $y \to 0$ and large $|x|$. The cross-like pattern persistent at large distances implies that the vortex trajectories propagate along the $x$ and $y$ axis only but not along, say, a diagonal (on average) direction in the $xy$ plane. Therefore, the rotational symmetry is strongly broken in the other phase,  $\beta_\tau > \beta_{\tau,c}$, with vortices forming spatially ordered structures along the $x$ and $y$ axes. What do these vortex structures look like?

\begin{figure}[t]
\centering
  \includegraphics[width=0.99\linewidth]{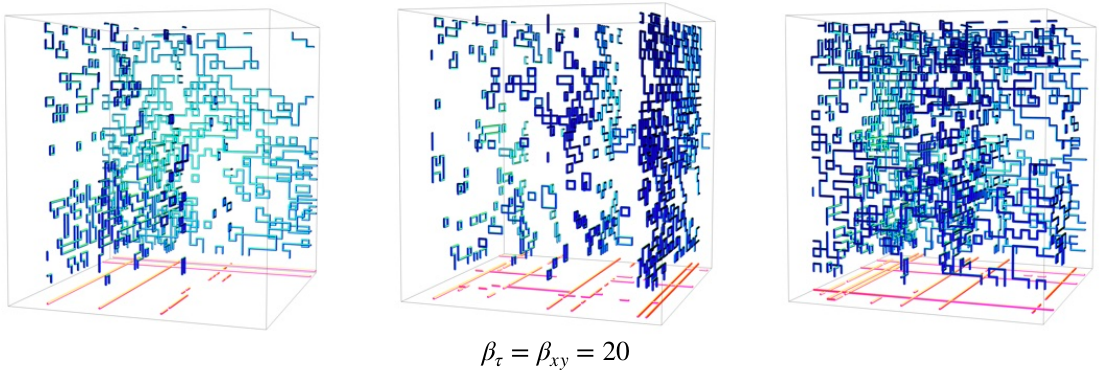}
  \caption{Three typical vortex configurations in a vortex wall phase at $\beta_\tau = \beta_{xy} = 20$. The vortex trajectories are shown in the blue/cyan colors, and their projections on the bottom $xy$ plane are presented in the pink/yellow colors.}
  \label{fig_configurations}
\end{figure}

\subsection{The vortex wall phase}

Typical examples of three-dimensional vortex configurations in a deep region of the $\beta_\tau > \beta_{\tau,c}$ phase are displayed in Figs.~\ref{fig_configurations}. Since the vorticity is a conserved quantum number~\eqref{eq_dual_conservation}, the vortex trajectories are closed lines. The configurations shown in Figs.~\ref{fig_configurations} clearly show the nature of the vortex anisotropy: the vortices are clustered into several well-separated walls parallel to either the $x$ axis or the $y$ axis. The vortex walls are thin, flat structures with thicknesses of about one or two lattice spacings. Thus, even visually, the three-dimensional spacetime vortex configurations have an evident signature of fractonic two-dimensional spacetime structures, the presence of which is consistent with the anisotropic property of the vortex correlations shown in the points D, E, and F of Fig.~\ref{fig_correlations}(a). According to analytical estimations, topological defects with a restricted, fractonic mobility are expected to appear in a wide class of models, including the XY plaquette model~\eqref{eq_S}, following algebraic~\cite{You:2019cvs} as well as entropic~\cite{You2022} arguments.

Thus, a typical field configuration possesses a collection of thin vortex walls that can only intersect with each other at the right angle ($\pi/2$). The space between the walls stays in the ordered phase, which is free of a disorder that is brought by (or revealed via the presence of) the vortices. However, the dynamics of the vortices still has a certain degree of disorder: the vortex trajectory within each wall is a collection of a large number of closed vortex lines supplemented by a single but geometrically large percolating vortex cluster. The disorder is only confined to the two-dimensional spacetime structures, while the fields in between the walls are in the ordered phase. 

Notice that the closed loops represent vortex-antivortex pairs that are first created from the vacuum by quantum fluctuations and then annihilate quickly, leaving a vortex-less vacuum after the annihilation. On the contrary, the large percolation loops demonstrate the persistent presence of vortices and anti-vortices in the ensemble. 

The appearance of the static vortex walls at large values of the lattice couplings $\beta_{xy}$ and $\beta_\tau$ can qualitatively be understood as follows. As the couplings become large, the partition function gets saturated by classical spin configurations that minimize action~\eqref{eq_S}, supplemented by ultraviolet fluctuations of the spin fields around these classical configurations. Quantum fluctuations lead to a pre-factor in a partition function that may either suppress or enhance the weight of a classical configuration. It is important to stress that the configurations are not fully classical but rather semiclassical (a classical background with quantum fluctuations around it). This property will play an essential role in the existence of the vortex walls. 

A large value of the time link coupling $\beta_\tau$ enforces the minimization of the link action~\eqref{eq_S_tau}, which implies that the spins separated by a single lattice spacing in the temporal direction $\hat \tau$ should coincide with each other:
\begin{align}
   (\Delta_\tau \phi)_{\boldsymbol{x},\tau} & = 2 \pi n_{\boldsymbol{x},\tau}, 
   \quad (n_{\boldsymbol{x},\tau} \in {\mathbb Z}), \nonumber\\
   & \Rightarrow \quad
   \phi_{{\boldsymbol x} + \hat \tau,\tau} =  \phi_{{\boldsymbol x}, \tau}\,.
   \label{eq_tau_minimization}
\end{align}
Applying this rule recursively along the time-like direction, we arrive at the time-independence of the configurations of the spin fields. Thus, in a strict classical limit, the spin field must be a static function, $\phi_{\boldsymbol{x}} \equiv \phi(x,y,\tau) = \phi(x,y)$, which depends only on the spatial coordinates $x$ and $y$. Consequently, the vortices in a classical configuration would be static as well. 

A large value of the spatial plaquette coupling $\beta_{xy}$ implies that the plaquette term attains its classical value:
\begin{align}
	\Delta_x \Delta_y \phi_{\boldsymbol{x}} = 2 \pi n_{\boldsymbol{x},xy}, 
   \quad (n_{\boldsymbol{x},xy} \in {\mathbb Z})\,.
   \label{eq_xy_minimization}
\end{align}
The most general solution to this equation is 
\begin{align}
    \phi_{\boldsymbol{x}} \equiv \phi(x,y) = f(x) + g(y)\,, 
\label{eq_phi_generic}
\end{align}
where $f$ and $g$ are arbitrary functions. One could also have forced the field $\phi_{\boldsymbol{x}}$ to appear in the canonical interval of values, $- \pi < \phi_{{\boldsymbol{x}},\mu} \leqslant \pi$, but we omit this requirement as it is not essential in our considerations below. The solution is independent of the imaginary time $\tau$ to minimize the link term in the action. 

One of the examples of the configurations is given by the linear fields:
\begin{align}
    \phi(x,y,\tau) = \phi_0 + \frac{2 \pi n_x}{L_x} x + \frac{2 \pi n_y}{L_y} y\,, 
    \qquad n_{x},n_y \in {\mathbb Z}\,,
    \label{eq_phi_linear}
\end{align}
where $\phi_0$ is an arbitrary constant field, and the integer numbers $n_{x}$ and $n_y$ enforce the periodicity of the solution along the $x$ and $y$ axis: $\phi(x+L_x, y, \tau) = \phi(x, y+L_y, \tau) = \phi(x, y, \tau)$.

Equation~\eqref{eq_phi_linear} takes the form of a pure (large) gauge configuration~\eqref{eq_gauge_transformation} which has, however, a nontrivial topology if the winding numbers $n_x$ and $n_y$ are nonzero. The vortex currents~\eqref{eq_v_correspondence} correspond to the vortex plaquettes~\eqref{eq_v_P} which are, in turn, are generated by the nontrivial links~\eqref{eq_bar_phi}. These links appear when the scalar field crosses the value ${[\phi(x,y,\tau)]}_{2\pi} = \pi$ so that its value is slightly below (above) $\pi$ at the beginning (end) of the link. Then, bringing the lattice derivative to the canonical $(-\pi,+\pi]$ interval adds a $2\pi$ phase to the links~\eqref{eq_bar_phi}, which will, at the next step, generate a set of vortex plaquettes~\eqref{eq_v_P} around that link. 

For example, let us take, for simplicity, $n_x = 0$ and $\phi_0 = 0$ in the linear configuration~\eqref{eq_phi_linear}. For $n_y = +1$, the spin field $\phi = 2 \pi y / L_y$ will cross the $\pi$ value at the middle of the lattice. In detail, this configuration will generate a set of a large number, $L_x \times L_\tau$ single links that will form a ``$2\pi$ rim'' normal to the $y$ axis. The nonzero links in the rim are normal to the $x\tau$ plane located around the middle point $y = L_y/2$. For $n_y = +2$, there will be two similar rims located on the equally spaced planes at $y = L_y/3$ and $y = 2 L_y/3$, and so on. A non-zero value of $\phi_0$ would shift these rims along the $y$ axis. If we take an arbitrary function $g(y)$ instead of the linear function in $y$, then these $2\pi$ rims will appear non-equally spaced. A field configuration with $n_x \neq 0$ will generate the $2\pi$ rims in the $y\tau$ plane. These rims are precursors of the vortex planes seen in Fig.~\ref{fig_configurations}.

Due to the $2\pi$ periodicity that enters the definition of vortex line, Eqs.~\eqref{eq_v_P} and \eqref{eq_v_correspondence}, the physical links in the $y$ direction that cross the $2\pi$ rim will experience a $2\pi$ jump, implying that $l_{x,\hat y} = \pm 1$ across the rim. The dual plaquette, which lives on the dual lattice and pierces the $\{x,\hat y\}$ links of the original lattice, will take the same value according to Eq.~\eqref{eq_v_correspondence}. The vortices are defined as the boundaries of the dual plaquettes that carry nonzero values. A single $l_{x,\hat y} = \pm 1$ non-trivial link will create a short vortex loop that encompasses an elementary plaquette of the dual lattice. One can, however, observe -- taking as an example a $n_x = 0$ wall of Eq.~\eqref{eq_phi_linear} -- that this constriction will lead to a collection of parallel plaquettes that form a closed (via the periodic boundary conditions) surface. The closed dual surface has no boundary, thus implying that the discussed construction will not lead to any vortex trajectory. In other words, the classical configuration~\eqref{eq_phi_linear} will not possess vortices at all. 

However, as we discussed above, the configurations that saturate the partition function have a semiclassical nature. The positions of the planes, determined by the procedure described above, are very sensitive to small fluctuations of the fields around the threshold $\pi$ value. The sensitivity is a result of the $2\pi$ normalization, which distinguishes, for example, the values $2 \pi \times 1.01$ [the vortex plaquette~\eqref{eq_v_P} is singular] and $2 \pi \times 0.99$ with no singularity. One can, therefore, show that a small perturbation of the field $\phi$ may change the position of a singular link by one lattice spacing. The latter one will make the dual surface open (no more closed) by creating a pair of short vortex loops. 

Thus, a weak perturbative disorder can generate vortex sheets (walls) around these two-dimensional manifolds~\eqref{eq_v_P}. Since the perturbations will appear all along the whole flat surface of the semiclassical configuration, we arrive at a vortex wall made of small vortex loops. The ultraviolet loops will thus be generated along the entire $2\pi$ rim surface. These loops will necessarily intersect with each other, thus forming, among small vortex loops, a long percolating vortex loop that will serve as a physical essence of the vortex wall.

The vortex walls are static two-dimensional flat structures that span $x\tau$ or $y\tau$ planes. Within each vortex wall, the individual vortex trajectories form a complicated, apparently randomly organized net that features various processes experienced by the vortex ensemble. The vortex, moving along the line, can annihilate with an anti-vortex moving along the same line. Conversely, a vortex-antivortex pair can be created from the vacuum. Finally, two vortices may elastically scatter with their motion being confined to the same line. 

The vortex motion, up to small local fluctuations, has a mostly one-dimensional character. However, when the $x\tau$ and $y\tau$ planes intersect, then a vortex may turn from the line parallel to the $x$ axis to the other line, which aligns with the $y$ direction. 


Summarizing, we expect that in the fully disordered phase (to the left of the transition line), the vortices propagate in the whole spatial two-dimensional plane, while in the vortex-walled phase (to the right of the phase transition line), the vortex moves predominantly along a set of one-dimensional spatial lines. Apparently, the change in the vortex dynamics from the two-dimensional behavior to the one-dimensional regime\footnote{It is important to stress our wording about the dimensionalities: if the spatial motion of vortices has a one-dimensional character so that it is restricted to a line, then the vortex trajectories are confined to flat two-dimensional spacetime planes. Similarly, if the motion of vortices is not restricted at all, they move in the whole two-dimensional spatial plane, and their world trajectories occupy the whole three-dimensional spacetime volume.} appears to happen around the phase transition line.

\subsection{Fractal dimension of vortex trajectories}

In a geometrical sense, unrestricted two-dimensional motion can be distinguished from partially immobile one-dimensional propagation by computing the fractal dimension of the corresponding vortex trajectory. Fractal dimension measures how complex a shape of a curve is based on how the curve scales at different magnifications. Various definitions of fractal dimensionality (or similar quantities) characterize the dimension of the manifold occupied by the line~\cite{falconer2014fractal}.

One of the intuitively most straightforward definitions of the fractal dimension of a line is as follows. Let $L = L(R)$ be the length of a line trajectory enclosed in the sphere of the radius $R$. Then, the fractal dimension $D_f$ characterizes how rapidly the total length $L(R)$ rises with the increase of the radius $R$:
\begin{align}
L(R) \propto R^{D_f}\,, \qquad\ {\text{as}} \quad R \to \infty\,.
\label{eq_Df_def}
\end{align}
For a straight line, the dimension is simply one, $D_f = 1$. For a line which randomly goes over the whole space of the dimension $D$, the fractal dimension is maximal, $D = D_f$. The limit of large radius $R$ in definition~\eqref{eq_Df_def} is motivated by physical applications where the line represents a trajectory of a physical object of a finite thickness $d$ so that the mathematical definition is not affected by particularities of the physical system as long as $R \gg d$. Even for infinitely thin vortex cores -- as in our case, where the vortices are single links -- the fractal dimension has a sense only in the limit when the sphere becomes sufficiently large with respect to the typical size of short-ranged (random) ultraviolet fluctuations. Thus, in the application to an ensemble of vortex trajectories, we require $R \gg 1$ in Eq.~\eqref{eq_Df_def}, where the lattice spacing (the length of a single link) is taken to be equal to one. On the other hand, the size of the sphere should not be too large in order to avoid the influence of periodic boundary conditions. Thus, in our calculations below, we take $1 \ll R \ll L$, which amounts, in practical applications, $R = 2 \dots 7$. 

There is also a nontrivial question about the definition of the vortex length $L$, which enters the definition~\eqref{eq_Df_def}. In a conventional disordered phase, a typical vortex ensemble contains a large number of disconnected short vortex loops and, typically, a single long vortex trajectory. The short vortex loops correspond to the ultraviolet fluctuations, while the large, percolating loop is associated with the presence of a condensate that fills the whole space. For any individual finite-sized loop, the definition in Eq.~\eqref{eq_Df_def} gives a vanishing fractal dimension~\eqref{eq_Df_def} because the vortex length $L(R)$ flattens, $L(R) \propto R^0$, if the radius $R$ becomes larger than the size of the vortex loop. It is, therefore, clear that only infinitely long trajectories can contribute to the fractal dimension. Consequently, in the definition of the fractal vortex dimension~\eqref{eq_Df_def}, we ignore these short ultraviolet vortices by selecting -- using a computer algorithm -- the most extended vortex trajectory per each vortex configuration.

Since we have found a single-phase transition, it is convenient to consider the properties of the vortices, such as the fractal dimension, along a selected line that crosses that transition line. A natural choice is given by the diagonal line $\beta = \beta_{xy} = \beta_{\tau}$ parameterized by the single parameter~$\beta$. At the diagonal line, the transition appears at the point:
\begin{align}
   \beta_{c} \equiv \beta_{\tau, c}(\beta_{xy} = \beta_{\tau}) = 1.00(7)\,,
   \label{eq_beta_c}
\end{align}
which separates the disordered phase at $\beta < \beta_c$ from the vortex-walled phase at $\beta > \beta_c$.

\begin{figure}[t]
\centering
  \includegraphics[width=0.9\linewidth]{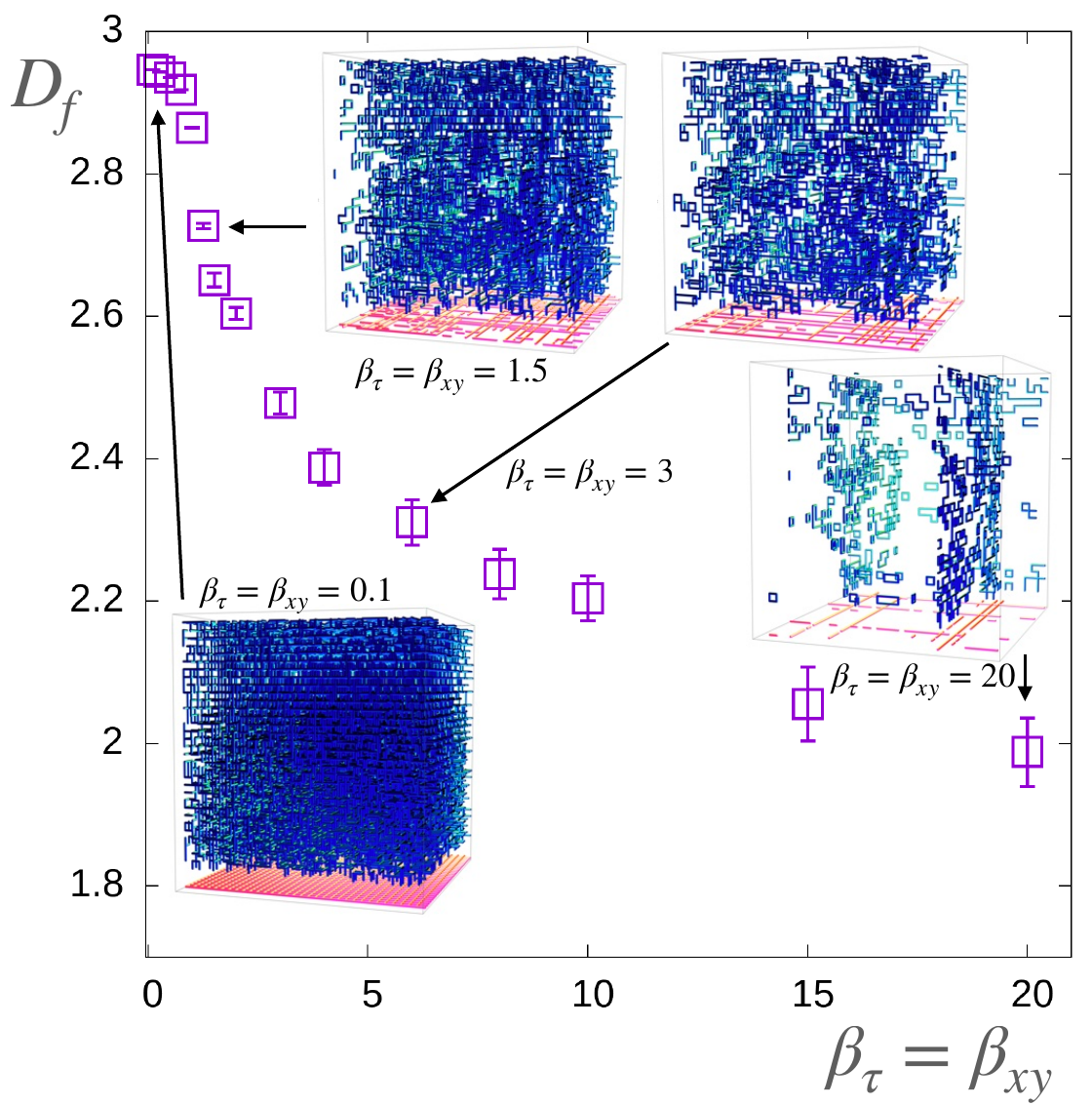}
  \caption{Fractal dimension $D_f$ of the vortex trajectories is shown along the diagonal line in the coupling $(\beta_\tau, \beta_{xy})$ space vs. the common lattice coupling $\beta = \beta_\tau = \beta_{xy}$. The insets display, at a selected set of values of the lattice couplings, typical vortex configurations with notation identical to Fig.~\ref{fig_configurations}.}
  \label{fig_Df_conf}
\end{figure}

In Fig.~\ref{fig_Df_conf}, we show the fractal dimension $D_f$, taken as an average over statistically significant configurations, as the function of the lattice coupling $\beta$ along the diagonal $\beta_\tau = \beta_{xy}$. This picture clearly illustrates the dimensional crossover from the totally disordered phase (where $D_f \simeq 3$) at low $\beta$ and the vortex wall phase (where $D_f \to 2$) at large $\beta$. In both cases, the vortex trajectories exhibit a disorder. They occupy the whole three-dimensional spacetime at the $\beta < \beta_c$ phase and get confined to the two-dimensional walls in the limit of the deep $\beta > \beta_c$ phase. In terms of the spatial dynamics, Fig.~\ref{fig_Df_conf} shows that while the vortices move freely along the whole two-dimensional plane at low $\beta$, their mobility gets partially restricted to static one-dimensional lines. The correlators shown in Fig.~\ref{fig_correlations}(a) clarify that these lines are parallel to either $x$ and $y$ axes. The insets in Fig.~\ref{fig_Df_conf} visually illustrate how the vortices experience the dimensional reduction from mobile to a partially immobile regime as the fractal dimension drops down with the increase of the ``diagonal'' lattice coupling~$\beta$.

\begin{figure}[t]
\centering
  \includegraphics[width=0.85\linewidth]{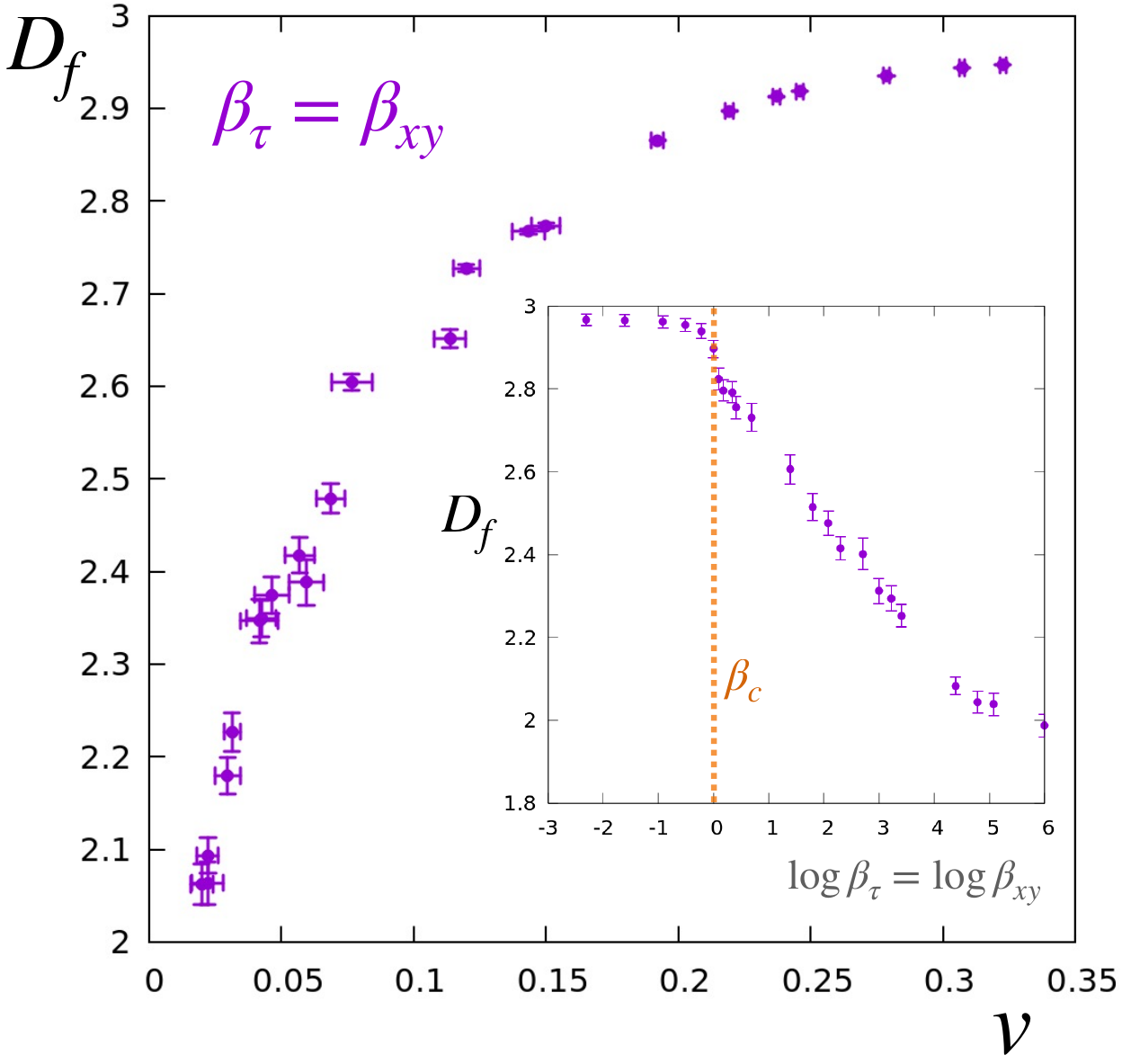}
  \caption{Parametric plot of the fractal dimension $D_f$ vs. the total vortex density $v$ for the lattice coupling $\beta_\tau = \beta_{xy}$ running along the diagonal line of the $(\beta_\tau, \beta_{xy})$ coupling plane. The inset shows the behavior of the fractal dimension $D_f$ along the same line (a common $\beta$ is shown in the logarithmic scale). The critical transition at the diagonal line, $\beta_c = \beta_{\tau,c} = \beta_{xy,c} \simeq 1.00(7)$, is shown by the orange dashed line in the inset figure.}
  \label{fig_Df_v}
\end{figure}

The main plot in Fig.~\ref{fig_Df_v} shows a curious correlation of the total vortex density,
\begin{align}
    v \equiv \frac{1}{3} \sum_{i=1}^3 \avr{|v_i|} = \frac{1}{3 L^3} \avr{\sum_{\boldsymbol{x}} \sum_{i>j=1}^3 \left| v_{\boldsymbol{x},ij} \right|}\,, 
\end{align}
(that counts both time-like and space-like vortex segments) with the fractal dimensionality of the vortex configurations~\eqref{eq_Df_def}. The lower bulk vortex density corresponds to the vortex-walled phase, which has the fractal dimension close to $D_f = 2$. On the contrary, the disordered phase possesses $D_f = 3$ at high vortex density as the vortices fill the whole space.

How fast does the change from the free propagation to the partially immobile vortex motion take place as we move from one phase to another? How does $D_f$ behave in the vicinity of the phase transition? To answer these questions, we show the behavior $D_f = D_f(\beta)$ in the logarithmic scale in the inset of  Fig.~\ref{fig_Df_conf}. The fractal dimension $D_f$  remains almost constant $D_f \simeq 3$ in the whole disordered phase. It starts to diminish only in the vicinity of the phase transition, at $\beta \approx \beta_c$. Exactly at the phase transition point, the dimension of the vortex lines drops down almost stepwise (while the magnitude of the step is small, it has a large slope). In the vortex-wall phase, the fractal dimension diminishes exponentially fast towards the partially immobile phase with $D_f \simeq 2$.

\corr{Notice that the effective theory of an exciton Bose liquid, considered in Ref.~\cite{Lake2022} reduces, following the renormalization group analysis, to the plaquette XY model~\eqref{eq_S}. Therefore, our resuts, that reveal the existence of the fractonic vortex phase transition in this spin model can potentially be relevant to the classical three-dimensional dimer-plaquette model which displays a fractonic BKT transition~\cite{Lake2022}.}

\section{Conclusions}

In conclusion, the XY-plaquette model offers a clear and accessible framework for representing a wide class of fractonic field theories, particularly those characterized by quasiparticles with limited mobility (fractons). The plaquette interaction appears natural as a ring-exchange term in various low-energy contexts of condensed matter systems such as exciton Bose liquids, cold atomic gases, and quantum dimer models. The model serves as a lattice counterpart of the scalar field model in the continuum limit~\eqref{eq_L_continuum_rescaled} which serves as the simplest prototype of a fractonic field model.

By employing first-principle Monte Carlo simulations, we analyzed the phase diagram and vortex dynamics of the XY-plaquette mode in two spatial dimensions on a square lattice. The minimal formulation of this model, given by Eq.~\eqref{eq_S}, possesses two phases associated with vortex dynamics. We found the expected disordered phase where the vortex trajectory fills in the entire spacetime and the partially disordered ``vortex-wall'' phase, where vortices exhibit constrained mobility. The critical line that separates these phases is characterized by the first-order and continuous transitions. \corr{Our findings of the nontrivial phase diagram appear to be in consistency with the early observation of the phase transition in a similar formulation of this model~\cite{Paramekanti2002}.}

In the disordered phase, the vortex trajectories possess the fractal spacetime dimension, $D_f \simeq 3$, while the second phase features $D_f \simeq 2$. In the vortex-wall phase, the motion of vortices is restricted to one or more straight lines (walls) parallel to the $x$ or $y$ axis. Within these lines, the individual vortices form a disordered system, leading to a fractal spacetime dimension for vortex trajectories approaching $D_f = 2$. The positions of the vortex walls emerge spontaneously, with $x$- and $y$-oriented lines intersecting at the right angles. We argued that the appearance of the vortex walls is a consequence of the spontaneous breaking of a global internal symmetry in the compact XY-plaquette model.

Thus, the compact version of the simplest field-fractonic model possesses topological vortex defects that behave as fractons (partially immobile excitations) themselves in one of the phases. 

\vspace{2ex}
\begin{acknowledgments}
AMB acknowledges generous support from the Carl Trygger Foundation Grant No. CTS 18:276. VAG and AVM were supported by Grant No. FZNS-2024-0002 of the Ministry of Science and Higher Education of Russia. AVM thanks the Institut Dennis Poisson (Tours, France) for the kind hospitality and acknowledges the support of Le Studium (Orl\'eans, France) research professorship. The work of MNC has been partially supported by the French National Agency for Research (ANR) within the project PROCURPHY ANR-23-CE30-0051-02.
\end{acknowledgments}

\bibliography{fractons}

\end{document}